\documentclass[aps,preprint]{revtex4}%
\usepackage{amsfonts}
\usepackage{amsmath}
\usepackage{amssymb}
\usepackage{graphicx}%
\begin{document}
\preprint{RBP03\_05/(CuInSe2ArXiv.tex)}
\title[CuInSe$_{2}$: distortions]{Influence of distortion on the electronic band structure of CuInSe$_{2}.$}
\author{H. Tototzintle-Huitle}
\affiliation{Departamento de F\'{\i}sica, CINVESTAV, M\'{e}xico}
\author{J. Ot\'{a}lora}
\affiliation{Departamento de F\'{\i}sica, Universidad Pedag\'{o}gica y Tecnol\'{o}gica de
Colombia, Tunja, Colombia.}
\author{J. A. Rodr\'{\i}guez}
\affiliation{Departamento de F\'{\i}sica, Universidad Nacional de Colombia, Bogot\'{a}, Colombia.}
\author{R. Baquero}
\affiliation{rbaquero@fis.cinvestav.mx}

\begin{abstract}
We present a tight-binding calculation of the influence of distorsion on the
bulk electronic structure of the chalcopyrite CuInSe$_{2}$. We calculate the
ideal case and then the effect of the inclusion of the distortions. We analyze
our results in detail and conclude from a comparison with other work that the
distortions must be included in the Hamiltonian to get a proper account of the
electronic band structure. We use our new Hamiltonian to study the effect that
both the tetragonal and the anionic distortion have on the (112) surface
electronic band structure. We find this effect non-negligible..

\end{abstract}
\volumeyear{year}
\volumenumber{number}
\issuenumber{number}
\eid{identifier}
\date[Date text]{date}
\received[Received text]{date}

\revised[Revised text]{date}

\accepted[Accepted text]{date}

\published[Published text]{date}

\startpage{1}
\endpage{102}
\maketitle

\section{ Introduction}

The quest for room temperature ferromagnetic semiconductors that can be
matched to conventional semiconductors resulted in an increasing interest in
A$^{II}$B$^{IV}$C$_{2}^{V}$ as well as in A$^{I}$B$^{III}$C$_{2}^{VI}$
chalcopyrites\cite{zunger04}. These materials are also interesting as
non-lineal optic devices, infrared photodetectors and in solar cell
applications with a high efficiency-to-cost ratio\cite{t9,t10,t11,t12,t13,t14}.

Chalcopyrites are tetragonal centered crystallographic structures with eight
atoms in the unit cell basis. Their spatial group is the D$_{2d}^{12}.$ The
location and identification of the eight atoms in the CuInSe$_{2}$ unit cell
is shown in Table IV below.

Chalcopyrites can deviate from their ideal symmetry in two ways. The $c/a$
ratio can be different from its ideal value 2 and also the anion which lies in
the middle of a tetrahedron can slide along the central axis. In this paper,
we present a detailed study of a tight-binding calculation that includes the
influence that both distortions have on the bulk electronic band structure of
CuInSe$_{2}$ . It is straightforward to generalize this work to calculate
their influence in the whole series of Cu-based chalcopyrites. The final
result is a set of Hamiltonians precise enough for the whole series that allow
to study different surfaces, monolayers and interfaces of these materials in
more complicated systems and situations of current interest with a very simple method.

\section{Some details on the method}

We use the tight-binding method \cite{slaterkoster} to calculate the
non-distorted as well as the distorted case. This method has been used before
to describe chalcopyrites successfully \cite{froyen,blom,jairoarbey,daniel} as
well as the related zincblende semiconductor compounds \cite{daniel1,daniel2}.
A$^{I}$B$^{III}$C$_{2}^{VI}$

In the tight-binding method one constructs Bloch functions $\phi_{\upsilon
}^{\mu}(\mathbf{k,r})$ to describe an electronic orbital $\upsilon$ centered
at the position $\mathbf{\tau}+\mathbf{d}_{\mu}$ of the ion labelled by $\mu,$
as a linear combination of atomic-like orbitals $\psi_{\upsilon}^{\mu
}(\mathbf{r})$ \cite{slaterkoster}%

\begin{equation}
\phi_{\upsilon}^{\mu}(\mathbf{k,}r)=\frac{1}{\sqrt{N}}\sum_{\tau
}e^{i\mathbf{k.(\tau}+\mathbf{d}_{\mu})}\psi_{\nu}^{\mu}(\mathbf{r-(\tau
}+\mathbf{d}_{\mu}))
\end{equation}
where \textbf{k} is a Bloch vector in the First Brillouin Zone and N the
number of unit cells in the crystal volume considered. We describe the group
III metal In and the group IV anion with atomic-like orbitals of s- and
p$^{3}-$ symmetry. For Cu we consider a full s, p$^{3}$, d$^{5}$ basis. The
spin-orbit interaction is responsible for a crystal field splitting of the
heavy and light hole bands on the top of the valence band. It is not taken
into account in our calculation since it is expected to be small. The matrix
elements of the Hamiltonian have the form:

\begin{center}%
\begin{equation}
\int\phi_{\upsilon}^{\mu}(\mathbf{k,r})\mathit{H}\phi_{\upsilon^{\prime}}%
^{\mu^{\prime}}(\mathbf{k}^{\prime}\mathbf{,r}^{\prime})d\mathbf{r=\delta
}_{\mathbf{k,k}^{\prime}}\sum_{\mathbf{d}_{\mu,\mu^{\prime}}}e^{i\mathbf{k.d}%
_{\mu,\mu^{\prime}}}<\nu|\nu^{\prime}>_{\mathbf{d}_{\mu,\mu^{\prime}}}%
\end{equation}

\end{center}

\bigskip where

\begin{center}%
\begin{equation}
<\nu|\nu^{\prime}>_{\mathbf{d}_{\mu,\mu^{\prime}}}=\int\psi_{\nu}^{\mu\ast
}(\mathbf{r})\mathit{H}\psi_{\nu^{\prime}}^{\mu^{\prime}}(\mathbf{r-d}_{\mu
\mu^{\prime}})d\mathbf{r\equiv}V_{\nu\nu^{\prime}}^{\mu\mu^{\prime}}
\label{elementos}%
\end{equation}

\end{center}

$\mathbf{d}_{\mu\mu^{\prime}}$ is the position vector of the $\mu^{\prime}$
atom from the $\mu$ atom. To calculate the non-diagonal matrix elements in
\ref{elementos} we use the Harrison's rule \cite{harrison}. Therefore the
interaction between an atomic-like orbital of symmetry $x$ located at the site
$\mu=1$ (In) with another atom of symmetry $y$ at $\mu^{\prime}=2$ (Se) is
given by $V_{xy}^{12}=lm[V(pp\sigma)-V(pp\pi)]$ . To actually calculate the
tight-binding parameters, we use further $V(ij\alpha)=\eta(ij\alpha)\hbar
^{2}/md_{\mu\mu^{\prime}}^{2}$($d_{\mu\mu^{\prime}}$ is the interatomic
distance, $m$ the electron bare mass) for s and p atomic-like orbitals. For
the interaction between s, p with d orbitals, we use instead $V(id\alpha
)=\eta(id\alpha)\hbar^{2}r_{d}^{3/2}/md_{\mu\mu^{\prime}}^{7/2}$ (for $r_{d}$
see Table III and ref. \cite{harrison}). The $\eta(tw\alpha)$ parameters are
given in \cite{harrison}. If we go on and calculate the diagonal matrix
elements using the same procedure, we get an inadmissible large value for the
gap. If we try the tight-binding parameters proposed by Papaconstantopoulos
\cite{papa} for Cu metal, we do not get the right gap as well. Also the Cu
on-site parameters that reproduce correctly the electronic band structure of
the superconducting perovskite. YBa$_{2}$Cu$_{3}$O$_{7}$ fail. Cu orbitals
have an important influence on the gap edges in the electronic band structure
of the Cu-based chalcopyrites.

In the semiconducting Cu-based chalcopyrites, the s-like orbital plays a mayor
role in fixing the lower edge of the conduction band while the p-like one
influences the position of upper edge of the valence band. The d-like
Cu-orbital mostly fixes the value of the chalcopyrite gap. These orbitals and
the p-like Se ones repel each other and push the upper valence band edge
upwards so that the gap is diminished \cite{zunger1,zunger2,yodee}.
Consequently, we have fixed the Cu on-site parameters for the whole series in
such a way that we get the lowest possible deviation from the semiconducting
gap. More exactly, we have selected the three Cu on-site parameters so that
$\sum_{\operatorname{series}}(E_{\operatorname{series}}%
-E_{g\operatorname{series}}^{E})^{2}$ as a function of
$E_{\operatorname{series}}$ is minimum. $E_{g\operatorname{series}}^{E}$ are
the experimental values of the gap. Small further adjustments of the anion
(Se) p on-site parameter for each chalcopyrite allowed us to get the right
experimental gap for the whole series. The experimental values that we used
are quoted in the next Table I.

\begin{center}%
\begin{tabular}
[c]{|l|l|}\hline
Chalcopyrite & E$_{g}^{E}[eV]$\\\hline
\textit{CuAlS}$_{2}$ & 3.49\\\hline
\textit{CuAlSe}$_{2}$ & 2.67\\\hline
\textit{CuAlTe}$_{2}$ & 2.06\\\hline
\textit{CuGaS}$_{2}$ & 2.43\\\hline
\textit{CuGaSe}$_{2}$ & 1.68\\\hline
\textit{CuGaTe}$_{2}$ & 1.23\\\hline
\textit{CuInS}$_{2}$ & 1.53\\\hline
\textit{CuInSe}$_{2}$ & 1.04\\\hline
\textit{CuInTe}$_{2}$ & 1.02\\\hline
\end{tabular}

Table I- Experimental optical gap for the whole series of Cu-based
chalcopyrites considered to set the tight-binding parameters \cite{zunger1}.
\end{center}

The on-site tight-binding parameters that we get in this way are compared in
Table II to the ones obtained from the Harrison's formulas \cite{harrison}.

\begin{center}%
\begin{tabular}
[c]{|l|l|l|}\hline
Parameter & Harrison & This work\\\hline
E$_{s}[eV]$ & -6.92 & -14.55\\\hline
E$_{p}[eV]$ & -1.83 & -2.22\\\hline
E$_{d}[eV]$ & -20.14 & -16.97\\\hline
r$_{\mathbf{d}}[\mathring{A}]$ & 0.67 & 1.15\\\hline
\end{tabular}

Table II- The Cu on-site tight-binding parameters. The parameter
r$_{\mathbf{d}} $ is defined in \cite{harrison}.
\end{center}

\section{The Hamiltonian}

The Hamiltonian is labelled with the atom numbers in the following way. We
take into account first nearest neighbors interactions only. The Hamiltonian
matrix takes the form

\begin{center}%
\begin{equation}%
\begin{tabular}
[c]{|c|c|c|c|c|c|c|c|c|}\hline
& $In1$ & $Cu3$ & $Cu5$ & $In7$ & $Se2$ & $Se4$ & $Se6$ & $Se8$\\\hline
$In1$ & $H_{11}$ & $0$ & $0$ & $0$ & $H_{12}$ & $H_{14}$ & $H_{16}$ & $H_{18}
$\\\hline
$Cu3$ & $0$ & $H_{33}$ & $0$ & $0$ & $H_{23}^{+}$ & $H_{34}$ & $H_{36}$ &
$H_{38}$\\\hline
$Cu5$ & $0$ & $0$ & $H_{55}$ & $0$ & $H_{25}^{+}$ & $H_{45}^{+}$ & $H_{56}$ &
$H_{58}$\\\hline
$In7$ & $0$ & $0$ & $0$ & $H_{77}$ & $H_{27}^{+}$ & $H_{47}^{+}$ & $H_{67}%
^{+}$ & $H_{78}$\\\hline
$Se2$ & $H_{12}^{+}$ & $H_{23}$ & $H_{25}$ & $H_{27}$ & $H_{22}$ & $0$ & $0$ &
$0$\\\hline
$Se4$ & $H_{14}^{+}$ & $H_{34}^{+}$ & $H_{45}$ & $H_{47}$ & $0$ & $H_{44}$ &
$0$ & $0$\\\hline
$Se6$ & $H_{16}^{+}$ & $H_{36}^{+}$ & $H_{56}^{+}$ & $H_{67}$ & $0$ & $0$ &
$H_{66}$ & $0$\\\hline
$Se8$ & $H_{18}^{+}$ & $H_{38}^{+}$ & $H_{58}^{+}$ & $H_{78}^{+}$ & $0$ & $0$
& $0$ & $H_{88}$\\\hline
\end{tabular}
\end{equation}

\end{center}

The diagonal sub-matrices are 9x9 for Cu and 4x4 for In and Se. The
Hamiltonian matrix is altogether 42x42.

Obviously, H$_{33}$ = H$_{55}.$ These refer to Cu. H$_{11}$ = H$_{77}$ , they
refer to In. H$_{22}$ = H$_{44}$ = H$_{66}$ = H$_{88}$ which describe the Se
atoms. The non-diagonal sub-matrices describe the first-nearest neighbors
interactions. Their tight-binding parameters were computed from the Harrison's
formulas \cite{harrison}. The anion p-on-site parameter was adjusted further
to get the exact experimental value (see below) With these data, the
Hamiltonian can be built up straightforwardly \cite{hamil,tesis,tesishugo}.

\section{Results}

\subsection{The ideal case}

To get the experimental value for the gap (see Table II) in our calculated
band structure we did a small (about 8\%) further adjustment to the p-on-site
parameter for the Se atom. The Harrison formula gives 9.53 eV to be compared
with our 8.789 eV. The electronic band structure is presented in the next Fig.
1. There are 2 Cu, 2 In and 4 Se atoms in the unit cell. Each Se atom
contributes with 2, each In with 3 and each Cu with 1s+5d occupied electronic
states and therefore 26 bands in the valence band. The rest of the 42 bands
(16) appear as empty conduction bands. The band structure appears in the next
Fig. 1.

FIGURE 1

Both the top valence band and the bottom conduction band are approximately
parabolic at $\Gamma$ and therefore in some calculations the effective band
approximation should be a good one. The semiconducting optical gap is direct
and is calculated as the difference between the energies $\Gamma_{1c}$ and
$\Gamma_{4\nu}^{(2)}$ which is equal to 1.04 eV (a fitted value to the
experimental one). The matrix element for dipolar transitions between these
states,
%TCIMACRO{\TEXTsymbol{<}}%
%BeginExpansion
$<$%
%EndExpansion
$\Gamma_{4\nu}^{(2)}|\mathbf{r}|\Gamma_{1c}$%
%TCIMACRO{\TEXTsymbol{>} }%
%BeginExpansion
$>$
%EndExpansion
is different from zero along the z-axis (since z is an element of the
$\Gamma_{4}$ representation; it is proportional to
%TCIMACRO{\TEXTsymbol{<}}%
%BeginExpansion
$<$%
%EndExpansion
$\Gamma_{4\nu}^{(2)}|\Gamma_{4}|\Gamma_{1c}$%
%TCIMACRO{\TEXTsymbol{>} }%
%BeginExpansion
$>$
%EndExpansion
and $\Gamma_{4}\otimes\Gamma_{1}=\Gamma_{4})$ but equal to zero at $\Gamma$
(since
%TCIMACRO{\TEXTsymbol{<}}%
%BeginExpansion
$<$%
%EndExpansion
$\Gamma_{4\nu}^{(2)}|z|\Gamma_{4\nu}^{(2)}$%
%TCIMACRO{\TEXTsymbol{>}}%
%BeginExpansion
$>$%
%EndExpansion
$\Longrightarrow$%
%TCIMACRO{\TEXTsymbol{<}}%
%BeginExpansion
$<$%
%EndExpansion
$\Gamma_{4}|\Gamma_{4}|\Gamma_{4}$%
%TCIMACRO{\TEXTsymbol{>} }%
%BeginExpansion
$>$
%EndExpansion
and $\Gamma_{4}\otimes\Gamma_{4}=\Gamma_{1}).$

\subsubsection{The valence band}

Immediately below the singlet state $\Gamma_{4\nu}^{(2)}$ (at the top of the
valence band), we find a doublet $\Gamma_{5\nu}^{(2)}$. The dipolar moment
along z is different from zero (%
%TCIMACRO{\TEXTsymbol{<}}%
%BeginExpansion
$<$%
%EndExpansion
$\Gamma_{5\nu}|z|\Gamma_{5\nu}$%
%TCIMACRO{\TEXTsymbol{>} }%
%BeginExpansion
$>$
%EndExpansion
$\Longrightarrow$ $<\Gamma_{5}|\Gamma_{4}|\Gamma_{5}$%
%TCIMACRO{\TEXTsymbol{>} }%
%BeginExpansion
$>$
%EndExpansion
and $\Gamma_{4}\otimes\Gamma_{5}=\Gamma_{5}).$ Therefore the dipolar moment at
the top of the valence band (a triplet in the zincblende parent compound)
breaks into a zero dipolar moment at the top $\Gamma_{4\nu}^{(2)}$ and a
non-zero one at the doublet $\Gamma_{5\nu}^{(2)}.$ The operator representing
the quadrupole moment is proportional to 3z$^{2}-r^{2}$ which belongs to
$\Gamma_{1}$ and the products of the type
%TCIMACRO{\TEXTsymbol{<}}%
%BeginExpansion
$<$%
%EndExpansion
$\Gamma_{x}|\Gamma_{1}|\Gamma_{x}$%
%TCIMACRO{\TEXTsymbol{>} }%
%BeginExpansion
$>$
%EndExpansion
are always different from zero since $\Gamma_{1}\otimes\Gamma_{x}=\Gamma_{x}$
and so a non-zero quadrupole moment will exist for all the valence band states.

The 26 bands that conform the valence band are grouped together into three
sub-bands separated by two in-band gaps. The first one (A in Fig.1) separates
the upper valence band (UVB) from the middle valence band (MVB) and the
in-band gap B separates this band from the lower valence band (LVB), At the
top of the UVB there is a singlet, $\Gamma_{4\nu}^{(2)},$ separated from a
doublet, $\Gamma_{5\nu}^{(2)},$ by a crystal field splitting, $\Delta_{cfs}$,
of 16 meV which is zero in the parent zincblende compound as we mentioned
above. Notice that the doublet remains such from $\Gamma-Z$ but splits from
$\Gamma-X.$

The chalcopyrite crystal field brakes the zincblende symmetry in several ways.
First, there are two different cations instead of one which transforms the
symmetry from cubic to tetragonal. Secondly, the anion can be found displaced
along the center line of the tetrahedron that it forms together with the two
different cations. But also the tetragonal symmetry ($\frac{c}{2a}=1$ where
$c$ and $a$ are the lattice parameters) is broken. We will deal with the
effect of these distortions below.

\paragraph{The upper valence band (UVB)}

The splitting of the triplet in the zincblendes into the singlet $\Gamma
_{4\nu}^{(2)}$ and the doublet $\Gamma_{5\nu}^{(2)}$ here, can be traced
easily to be due to the presence a the two cations. This can be done with the
tight-binding program replacing the parameters for Cu with those of In to get
the bands for InSe or reversing the way we replace the parameters, we can get
CuSe. In both cases the bands show a triplet on the top of the valence band at
$\Gamma.$When the bands of the zincblende are compared to the ones of the
chalcopyrite, we realize that two further splitting occur at the top of the
valence band, one at Z ($\Delta Z$ in Fig. 1) and another one at X ($\Delta
X).$

There are 10 bands in the UVB that lie from 0 to -5 eV (the origin is set at
the top of the valence band in $\Gamma$ as it is customary). The main
contribution comes from p-like Se orbitals. The details of the composition are
in Fig.2 where the density of states (DOS) is shown. The shadow areas are
proportional to the contribution of the orbital identified at the upper right corner.

\paragraph{ The middle valence band (MVB)}

The inner-band gap A is small (16 meV, see Figs. 1 and 2). The MVB contains 12
bands; 10 of them are d-Cu orbital contributions. The deepest band of this
group runs from $Z_{4\nu}+Z_{5\nu}\rightarrow\Gamma_{4\nu}^{(1)}\rightarrow
X_{1\nu}^{(4)}$ as shown in Fig.1.

\paragraph{The lowest valence band (LVB)}

The deepest group of bands, the LVB, is separated from the MVB by a large gap
from $\Gamma_{4\nu}^{(1)}$ to $\Gamma_{3\nu}$ of about 4eV. The main
contribution comes from the s-Se orbitals (see Fig.2 for more details). The
upper band of this group is a singlet $\Gamma_{3\nu}$ followed very closely by
a doublet $\Gamma_{5\nu}^{(1)}$. In the zincblende parent compound these bands
are degenerate. This splitting is due to the presence of a second cation. The
upper band of this group $Z_{1\nu}+Z_{2\nu}\rightarrow\Gamma_{3\nu}$ is doubly
degenerate from $Z-\Gamma$ but splits from $\Gamma-X.$

FIGURE 2

\subsubsection{The conduction band}

The conduction band (CB) minimum runs from $Z_{1c}+Z_{2c}\longrightarrow
\Gamma_{1c}\longrightarrow X_{1c}^{(1)}$ which is a singlet all along $Z\Gamma
X.$ At $Z$, nevertheless the band is degenerate and splits to a
higher-in-energy band that runs from $Z_{1c}+Z_{2c}\longrightarrow\Gamma
_{3c}\longrightarrow X_{1c}^{(1)}$. At $X$ the band is again degenerate but at
$\Gamma_{3c}$ it is a singlet. From Fig.2, we see that the CB is divided into
two clearly defined sub-bands separated by an in-band gap of about 0.2 eV.
Each sub-band presents two peaks. The DOS is quite higher in the upper part
the spectrum. The lowest conduction band (LCB) goes from roughly 1-5 eV. The
upper conduction band (UCB) runs from about 5.2-12 eV. The lower peak of the
LCB is composed mainly from s-Cu and s-Se orbitals in the 1-3.7 eV energy
region and from s-In and s-Se in the higher energy region. In the low energy
region of the UCB, the main contribution is from p-Cu orbitals and in the
higher energy peak it is from p-In ones (see Fig.2 for more details).

\subsubsection{Comparison with other work}

Jaffe and Zunger (JZ) \cite{zunger1} made an \textit{ab initio }calculation of
the electronic band structure of the chalcopyrite CuInSe$_{2}$. Sometimes ab
initio calculations do not get the semiconducting gap right. JZ got 0.98 eV.
The experimental value is 1.04 eV. But otherwise, our calculation compares
well with theirs. The valence band width ($\Gamma_{4\nu}^{(2)}-\Gamma_{\nu
}^{(1)})$ is 14.8 eV in our work and 13.8 eV in theirs. Both calculations
agree in the fact that the top of the VB which is a triplet in the zincblende
parent crystal structure is split apart into a singlet at the top and a deeper
doublet. The $\Delta_{cfs}=-0.08$ eV in JZ and -0.016 in this work. We will
show below that the difference shrinks as an effect of the distortions (both
anion and tetragonal which are considered by JZ). Yoodee \textit{et al.}
\cite{yoodee} has calculated this crystal field splitting $\Delta_{cfs}=0$ for
the ideal case and $\Delta_{cfs}=+0.01$ eV when the tetragonal distortion
($\frac{c}{a}=2.008)$ is taken into account. In his case the bands are in a
reverted order which means that the doublet is on the top of the valence band.
This can be associated to his neglect of the anion distortion. There are some
differences in both calculations. For example, for the width of the UVB we get
5 eV while JZ get 4 eV. This is actually the origin for the difference in the
overall VB width. It is worth mentioning that the in-band gap A differs
substantially in both works even qualitatively. The two bands that define this
gap are reverted in JZ's work giving a -0.01 eV while we get a broad in-band
gap of 1.67 eV. The overall width of the MVB does not differ very much. We get
1.5 eV and JZ 1.37 eV. On the other hand, the in-band gap B is higher in JZ
(7.39 eV) while we get 4.79. Some of the differences are shown in detail in
Table V below.

\subsection{Influence of distortions}

It os obvious that a correct description of the chalcopyrites must include the
effect of distortions (both, the anion and the tetragonal one). In this
section we will show the effect of this inclusion. The tetragonal distortion
means that the ratio $\frac{c}{2a}=\eta\neq1$ and the anion distortion means
that the anion is not located exactly at $a$($\frac{1}{2},\frac{1}{2},\frac
{1}{4}$) and equivalent positions but rather at $a$($u,\frac{1}{2},\frac{\eta
}{4}$). The new positions are shown in the next Table IV.

\begin{center}%
\begin{tabular}
[c]{|c|c|c|c|}\hline
Nb. & atom & ideal & distorted\\\hline
1 & In & $(0,0,0)$ & $(0,0,0)$\\\hline
2 & Se & ($\frac{1}{4},\frac{1}{4},\frac{1}{8})$ & $(u,\frac{1}{4},\frac{\eta
}{4})a$\\\hline
3 & Cu & ($\frac{1}{2},0,\frac{1}{4})$ & $(\frac{1}{2},0,\frac{\eta}{2}%
)a$\\\hline
4 & Se & ($\frac{3}{4},\frac{1}{4},\frac{3}{8})$ & $(\frac{3}{4},\frac{1}%
{2}-u,\frac{3\eta}{4})a$\\\hline
5 & Cu & $(0,0,\frac{1}{2})$ & $(\frac{1}{2},\frac{1}{2},0)a$\\\hline
6 & Se & ($\frac{1}{4},\frac{1}{4},\frac{5}{8})$ & $(1-u,\frac{3}{4}%
,\frac{\eta}{4})a$\\\hline
7 & In & ($\frac{1}{2},0,\frac{3}{4})$ & $(0,\frac{1}{2},\frac{\eta}{2}%
)a$\\\hline
8 & Se & ($\frac{3}{4},\frac{1}{4},\frac{7}{8})$ & $(\frac{1}{4},\frac{1}%
{2}+u,\frac{3\eta}{4})a$\\\hline
\end{tabular}

Table IV Positions of the atoms in the unit cell in the ideal and distorted case.
\end{center}

We use the same Hamiltonian described above with the same basis on each atom
and introduce the changes that describe the corresponding distortions. We can
use the same symmetry group to construct the Hamiltonian since X-ray
experiments indicate that the space group is preserved \cite{t10, t11,zunger1,
t7, t23, t24} in spite of the distortions present. In this case the Se-atom
(the anion) will have its first nearest neighbors (the cations) at different
distances. Namely,%

\begin{equation}
R_{In\longrightarrow Se}=R_{12}=a\sqrt{u^{2}+\frac{(1+\eta)^{2}}{16}}%
\end{equation}

\begin{equation}
R_{Se\longrightarrow Cu}=R_{23}=a\sqrt{(u-\frac{1}{2})^{2}+\frac{(1+\eta)^{2}%
}{16}}%
\end{equation}

and the direction cosines that intervene in the calculation are given by:

\begin{center}%
\begin{tabular}
[c]{|c|c|c|c|}\hline
Vector & l & m & n\\\hline
R$_{12}$ & $\frac{ua}{R_{12}}$ & $\frac{a}{4R_{12}}$ & $\frac{\eta a}{4R_{12}%
}$\\\hline
R$_{23}$ & $\frac{(\frac{1}{2}-u)}{R_{23}}$ & -$\frac{a}{4R_{23}}$ &
$\frac{\eta a}{4R_{23}}$\\\hline
\end{tabular}

Tabla IV The direction cosines once distortion is taken into account. These
are equal always to $\frac{1}{\sqrt{3}}$ in the ideal case.
\end{center}

The phase factors are to be corrected accordingly. Now it is straightforward
to construct the tight-binding Hamiltonian that includes both, the tetragonal
as well as the anion distortion. We present in the next Fig.3 our results for
the influence that the inclusion of distortion has on the electronic band
structure of CuInSe$_{2}.$

FIGURE 3

The general picture of the bands is quite similar. The VB present the same
three groups separated by two in-band gaps. We do not obtain an important
reduction of the total band width (13.79 eV to be compare to the ideal case
value of 13.8 eV). The conduction band is formed again by two groups separated
by a small in-band gap. Also as a general trend, we can say that in several
aspects the result including the distortions agrees better with the JZ calculation.

The first think to notice is that since the space group remains the same, the
degeneracies at the high symmetry points $Z$, $\Gamma$, and $X$ \ also remain
as it is to be expected. On the other hand, we see that the distortions have a
mayor or minor effect on almost all the bands. Distortion diminishes the
optical gap and as a consequence we get 0.95 eV in closer agreement with the
0.98 eV of JZ . The experimental value to which we have fitted the on-site
parameters in the ideal case is 1.04 eV. We could retouch the p-on site Se
parameter to get the experimental gap value back. We present our results here
without this further adjustment to leave sharp the picture of the effect of distortion.

In the UVB, the crystal field splitting $\Delta_{cfs}=-0.09eV$ (JZ got
$-0.08eV)$ in contrast to the ideal case value $-0.016eV.$ So the distortions
have a very important influence in this crystal field splitting. If we take
into account the result by Yoodee \textit{et al. }\cite{yoodee} quoted above,
then we can conclude that the most important factor that influences the
crystal field splitting is the anion distortion. The $\Delta Z$ and $\Delta X
$ splitting enhance due to distortions as can be seen from the enhancement of
the top valence band details in Fig.3. This sub-band broadens and as a
consequence the in-band gap A shrinks.

The next MVB appears to be still a very narrow band from the d-Cu orbitals
mainly. The effect of distortions is here a minor one and it is related
essentially to the slight shrinking of the total sub-band width which enhances
but slightly the in-band gap B. Notice that the band $Z_{4\nu}+Z_{5\nu
}\longrightarrow\Gamma_{4\nu}^{(1)}\longrightarrow X$ $_{1\nu}^{(4)}$ becomes
almost flat as a result of the distortions.

In the LVB which contains 4 bands associated with s-Se orbitals the main
effect is to invert the sign of the crystal field splitting which reverts the
order between $\Gamma_{3\nu}$ and $\Gamma_{5\nu}^{(1)}$ so that now the doubly
degenerate band is on top. We summarize the effect of distortion in Table VI
where we compare the energy values for several bands that we get for the ideal
case and for the result after distortions are included with the ones obtained
by JZ.

\begin{center}%
\begin{tabular}
[c]{|c|c|c|c|}\hline
State & JZ & Ideal & Distortioned\\\hline
UVB-maximum & (eV) & (eV) & (eV)\\\hline
$\Gamma_{4v}^{(2)}$ & 0..00 & 0.00 & 0.00\\\hline
$\Gamma_{5v}^{(2)}$ & -0.08 & -0.02 & -0,09\\\hline
$Z_{3v}+Z_{4v}$ & -0.79 & -1.93 & -1.82\\\hline
$X_{1v}^{(5)}$ & -0.54 & -1.83 & -1.22\\\hline
\multicolumn{4}{|c|}{UVB-minimum}\\\hline
$\Gamma_{4v}^{(1)}$ & -5.15 & -8.11 & -7.89\\\hline
$Z_{4v}+Z_{5v}$ & -5.12 & -7.98 & -7.87\\\hline
$X_{1v}^{(4)}$ & -5.13 & -7.92 & -7.82\\\hline
s-Se band &  &  & \\\hline
$\Gamma_{2v}^{(1)}$ & -13.03 & -12.91 & -12.82\\\hline
$\Gamma_{3v}$ & -13.06 & -12.90 & -12.90\\\hline
$\Gamma_{1v}^{(1)}$ & -13.83 & -14.80 & -14.79\\\hline
$Z_{1v}+Z_{2v}$ & -13.00 & -12.91 & -12.82\\\hline
$Z_{5v}$ & -13.46 & -14.00 & -13.99\\\hline
$X_{1v}^{(2)}$ & -13.20 & -13.36 & -13.38\\\hline
$X_{1v}^{(1)}$ & -13.31 & -13.57 & -13.47\\\hline
\multicolumn{4}{|c|}{Other values}\\\hline
Width band s-Se at $\Gamma$ & 0.80 & 1.90 & 1.96\\\hline
Gap A & -0.01 & 1.67 & 1.23\\\hline
Gap B & 7.39 & 4.79 & 4.94\\\hline
$\Delta z$ & 0.5 & 0.11 & 0.27\\\hline
$\Delta x$ & 0.4 & 0.26 & 0.37\\\hline
$\Gamma_{5v}^{(1)}-\Gamma_{3v}$ & 0.03 & -0.01 & 0.08\\\hline
\end{tabular}

Table V Comparation between energy values at some high simmetry points in the
ideal and distorted cases.
\end{center}

In the conduction band \ the major effect of distortion is to shrink the
sub-band width of the s-Cu orbitals and to enhance the one corresponding to
s-In orbitals. The in-band gap between the two enhances. We compare some CB
values in the ideal case with the ones when the distortions are included in
the calculation to show the effect of distortion and quote the corresponding
values obtained by JZ in the next Table VI.

\begin{center}%
\begin{tabular}
[c]{|c|c|c|c|}\hline
State & JZ & Ideal & Distorted\\\hline
Conduction band & (eV) & (eV) & (eV)\\\hline
$\Gamma_{1c}=E_{g}(gap)$ & 0.98 & 1.04 & 0.95\\\hline
$\Gamma_{3c}$ & 3.24 & 3.77 & 3.25\\\hline
$Z_{1c}+Z_{2c}$ & 2.76 & 3.14 & 2.83\\\hline
$X_{1c}^{(1)}$ & 2.25 & 1.87 & 1.56\\\hline
\end{tabular}

Table VI The effect of distortion in the Conduction Band.
\end{center}

\subsection{The (112) surface}

In the next Fig.4, we present our result for the effect that the distortions
have on the electronic band structure of the (112) surface of CuInSe$_{2}.$
The open circles are the ideal case and the black ones are the result when
distortions are included. In general, distortions do affect the VB and the CB
in a noticeable way.

FIGURE 4%

%TCIMACRO{\FRAME{ftbpFUX}{8.4877cm}{10.8601cm}{0pt}{\Qcb{\ Influence of
%distortions in the electronic band structure of the (112) surface of
%CuInSe$_{2}.$ Both, the tetragonal and the anion distortion are included. The
%open circles are the ideal result and the black ones the one with distortions
%included.}}{\Qlb{figure4}}{fig4.wmf}{\special{ language "Scientific Word";
%type "GRAPHIC";  maintain-aspect-ratio TRUE;  display "USEDEF";
%valid_file "F";  width 8.4877cm;  height 10.8601cm;  depth 0pt;
%original-width 6.0312in;  original-height 7.7323in;  cropleft "0";
%croptop "0.9988";  cropright "0.9993";  cropbottom "0";
%filename 'prbmar05/FIG4.wmf';file-properties "NPEU";}}}%
%BeginExpansion
\begin{figure}
[ptb]
\begin{center}
\fbox{\includegraphics[
trim=0.000000in 0.000000in 0.004222in 0.009279in,
natheight=7.732300in,
natwidth=6.031200in,
height=10.8601cm,
width=8.4877cm
]%
{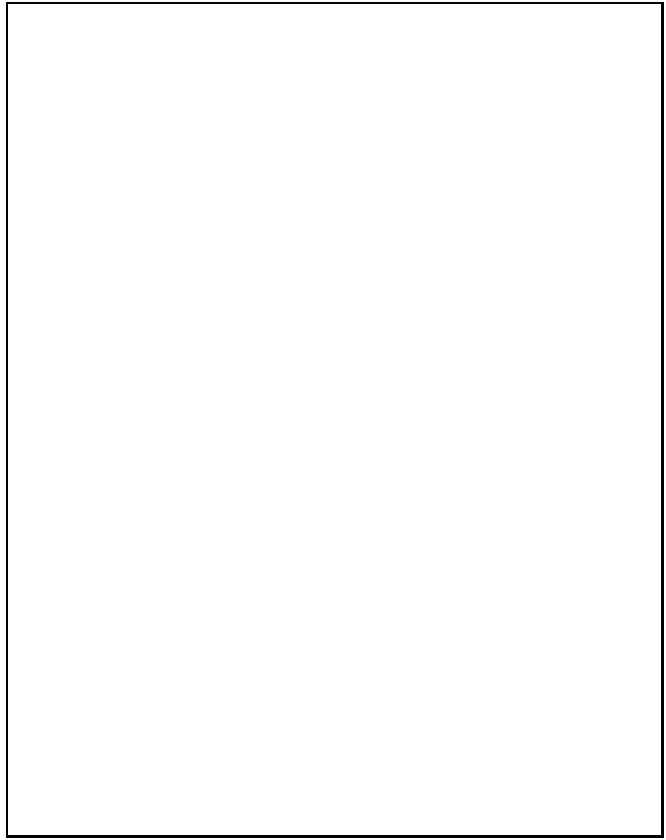}%
}\caption{\ Influence of distortions in the electronic band structure of the
(112) surface of CuInSe$_{2}.$ Both, the tetragonal and the anion distortion
are included. The open circles are the ideal result and the black ones the one
with distortions included.}%
\label{figure4}%
\end{center}
\end{figure}
%EndExpansion

Of great interests the effect of distortions on the surface states. The full
line in Fig.4 represents the border of the bulk valence band (VB) and of the
conduction band (CB). Three surface states denoted as 1SE, 2SE and 3SE in
Fig.4 appear. The 1SE state starts at $\Gamma$ somewhere in the middle of the
semiconducting gap. Both 2SE and 3SE appear to be closer to the CB minimum.
Somewhere in the interval \textit{K-J }the 2SE state crosses the state 1SE and
becomes the one nearest to the top of the VB. If we follow 1SE further on, we
see an important influence of distortions on this state. Indeed, in the ideal
case, 1SE enters the CB region and becomes a resonance while if distortions
are included, the state remains within the energy gap region as a pure surface
state. A similar effect occurs with the 3SE state around \textit{K} which is a
resonance in the ideal case but becomes a pure surface state as an influence
of distortions. The 2SE state is switched towards the top of the VB when
distortions are included. So the (112) CuTnSe$_{2}$ surface shows three pure
surface states as a consequence of distortions that would partially become
resonances would the lattice remain ideal. We conclude that distortions have
an effect also in the surface states that is important and cannot be neglected.

\section{Conclusions}

We have studied the effect of distortion in the electronic band structure of
the chalcopyrite CuInSe$_{2}.$ We find that its effect is important and that
the inclusion of both the tetragonal and the ionic distortion are important to
get a proper description of the electronic bands. Once these are included we
find that that the tight-binding Hamiltonian gives an accurate enough result
to be useful for further calculations of surfaces, monolayers and interfaces
and more complicated system that include this material. We have used the
Hamiltonian together with the Green%
%TCIMACRO{\U{b4}}%
%BeginExpansion
\'{}%
%EndExpansion
s Function Matching Method to calculate the effect of distortion on the (112)
surface of this material. We find that distortions have an important effect on
the three surface states that we found on this surface which is used in
important technological applications.

\newpage

FIGURE CAPTIONS

Fig.1 The electronic band structure of CuInSe$_{2}$.

Fig.2 \ Contribution to the DOS from the different orbitals at different energies.

Fig.3 Influence of distortion on the electronic band structure of
CuInSe$_{2}.$ We used $u=0.224$ and $\eta=1.004$ (the ideal values are 0.25
and 1 respectively). The full line is the ideal case and the dot lines
represent the effect of distortion.

Fig.4 \ Influence of distortions in the electronic band structure of the (112)
surface of CuInSe$_{2}.$ Both, the tetragonal and the anion distortion are
included. The open circles are the ideal result and the black ones the one
with distortions included.

\bigskip

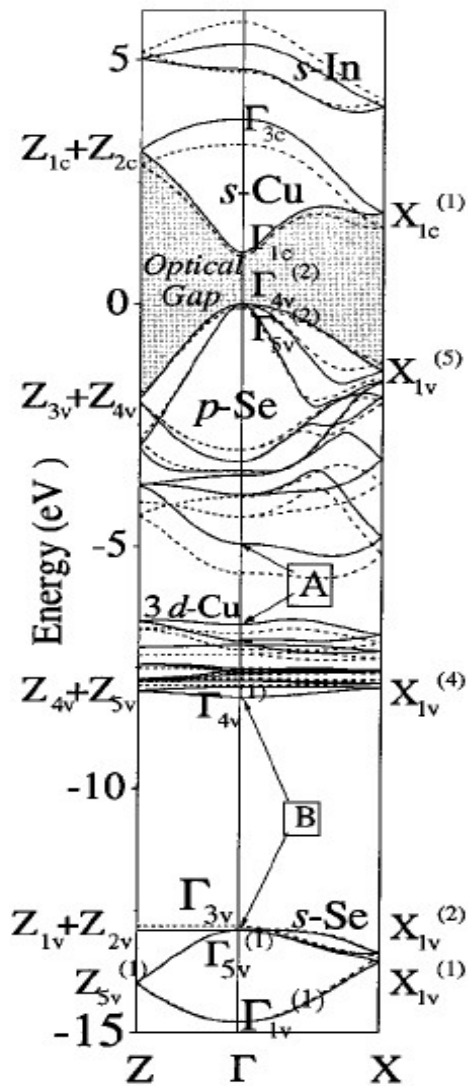
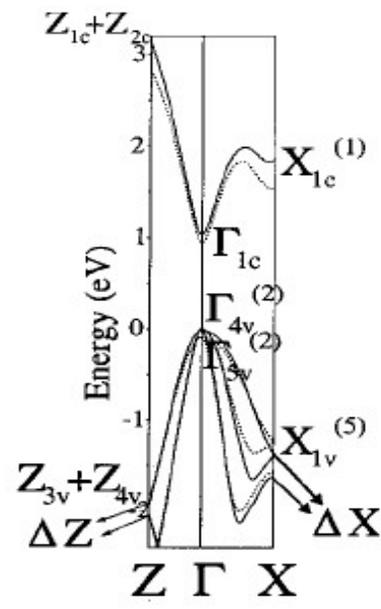
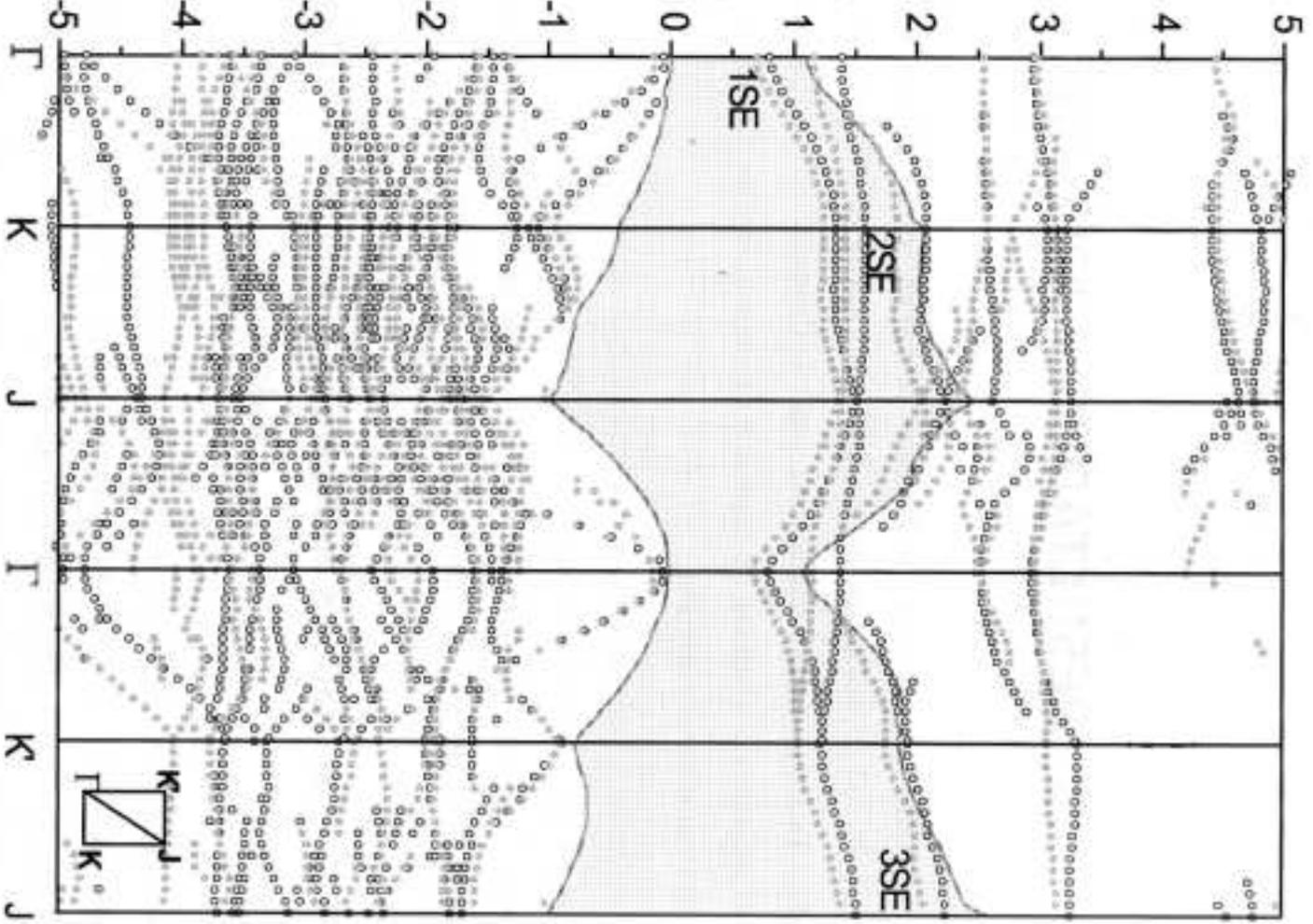
\end{document}